\documentclass[iop]{emulateapj} 
\usepackage{natbib}
\usepackage{graphicx}
\usepackage{amsmath}   

\graphicspath{{./}}

\newcommand{\Msun}{\rm M_{\odot}}

\newcommand{\MBH}{M_{\mathrm{BH}}}
\newcommand{\pc}{\rm {pc}}
\newcommand{\yr}{\rm {yr}}
\newcommand{\Mdotbh}{\dot{M}_{\rm BH}}
\newcommand{\Mdotedd}{\dot{M}_{\rm Edd}}

\newcommand{\cm}{\rm {cm}}
\newcommand{\ri}{r_{\rm ion}}
\newcommand{\rB}{r_{\rm B}}
\newcommand{\rs}{r_{\rm s}}

\begin{document}

\title{Accretion onto Intermediate-mass Seed Black Holes in Primordial Galaxies}

\author
{
Yuexing Li
}

\affil{Department of Astronomy \& Astrophysics, The Pennsylvania State University, 
525 Davey Lab, University Park, PA 16802, USA}

\email{yuexing@astro.psu.edu} 

\begin{abstract}

The origin of the supermassive black holes that power the most distant quasars observed is largely unknown. One hypothesis is that they grew rapidly from intermediate-mass seeds ($\sim 100\, \Msun$) left by the first stars. However, some previous studies argued that accretion onto these black holes was too low to build up the mass due to strong suppression by radiative feedback. Here, we re-exam the accretion process of such a black hole embedded in a primordial gas cloud, by considering a wide range of physical and numerical parameters not explored before. We find that, while radiative heating and pressure indeed suppress accretion effectively, self-gravity of the gas eventually overcomes the feedback effects and boosts the accretion to the Eddington rate after one free-fall timescale of the cloud. Moreover, for a given black hole mass, there exists a critical density above which the accretion can reach Eddington limit. Furthermore, we find a universal correlation between black hole accretion rate and ambient gas density, which may serve as a realistic recipe for black hole growth in simulations.

\end{abstract}

 
\section{INTRODUCTION}

A major recent development in observational cosmology has been the discovery of dozens luminous quasars at high redshifts ($z \gtrsim 6$) when the universe was less than 7\% of its current age \citep{Fan2001, Fan2003, Willott2007, Willott2010A, Mortlock2011}. These quasars are believed to be powered by supermassive black holes (SMBHs) of $\sim 10^{8-9}\, \Msun$, and the BHs appear to accrete at near Eddington rate (e.g., \citealt{Fan2006A, Willott2010B}). Furthermore, intense star formation \citep{Carilli2004, Walter2009}, abundant CO gas and dust \citep{Walter2003, Jiang2006, Jiang2010, Wang2010}, and super-solar metallicity \citep{Maiolino2005} were detected in the quasar hosts, indicating a co-eval formation of the SMBHs and host galaxies. 

The seeds and growth of these SMBHs, however, are unsolved puzzles. It has long been proposed that they might grow from remnants of the very first stars, the so-called PopIII stars (e.g., \citep{Madau2001, Haiman2001, Volonteri2005, Li2007A, Volonteri2010}). Sophisticated cosmological simulations over the past decade suggested that PopIII stars formed at early cosmic times $z \sim 20 - 50$ in minihalos with mass $\sim 10^{6}\, \Msun$, and were likely massive, with mass $\sim 30 - 300\, \Msun$ (e.g., \citealt{Abel2002B, Bromm2004, Tan2004, Gao2007A, Yoshida2008, Bromm2009}) (see, however, recent work by \citealt{Clark2011, Greif2011} for suggestions of a smaller mass range).  Massive PopIII stars of $100-140\, \Msun$ or above 260 $\Msun$ were predicted to rapidly collapse to BHs of $\sim 100\, \Msun$ \citep{Heger2003}. The subsequent growth of these stellar-mass BHs might play an important role in the formation of luminous quasars at later times. If a $100\, \Msun$ BH starts to accrete at the Eddington rate at $z>20$, it will attain a final mass greater than $10^9\, \Msun$ by $z \sim 6$, implying a viable explanation for the observations of bright quasars in this epoch. 

Black holes are thought to accrete through a disk shaped by the outward transfer of the angular momentum \citep{Shakura1973, Rees1984, Antonucci1993, Krolik1999}. However, in metal-free, primordial galaxies with virial temperature above $10^4$ K, conditions may exist for the formation of a thick disk of gas at temperature of 5000 - 10000 K due to insufficient cooling \citep{Oh2002, Volonteri2005}. As a result, accretion proceeds in a quasi-radial manner onto the central BH. Such spherical Bondi-Hoyle accretion model \citep{BondiHoyle1944, Bondi1952} has been widely employed in cosmological simulations on BH growth in the early universe \citep{Li2007A, Johnson2007, DiMatteo2008, Alvarez2009, Sijacki2009, DiMatteo2011}. In particular, it was shown by \cite{Li2007A} that $100\, \Msun$ BH seeds from PopIII stars in gas-rich protogalaxies residing in highly overdense regions could grow to $10^9\, \Msun$ within 800 million years and produce luminous quasars at $z \sim 6$ as those observed \citep{Li2008, Robertson2007, Narayanan2008}, provided that they accreted at Eddington rates for much of their early time. 

There are two primary feedback effects of the radiation from the BH that greatly influence the accretion process (e.g., \citealt{Ciotti2001, Sazonov2005, Ciotti2007, Johnson2007, Alvarez2009, Milos2009B, Milos2009A, Park2011}). The first one is the photo-ionization heating of the gas within the sonic radius (at which the gas motion becomes supersonic), which significantly increases the sound speed and reduces the accretion rate. The second one is the radiation pressure from both electron scattering and photo-ionization, which may drive an outflow and halt the accretion completely. To study these feedback processes and their effects on the accretion rate of the central BH, one must resolve the spatial scale smaller than the sonic radius, which is of the order of $10^{14}$~cm ($\sim 10^{-4}$~pc) for a 100~$\Msun$ BH. This clearly poses a significant challenge for large-scale cosmological simulations. 

Recent analytical \citep{Milos2009B} and idealized simulations \citep{Milos2009A, Park2011} of accretion onto a $100\, \Msun$ BH in a gas sphere argued that radiation feedback strongly suppressed the accretion rate to be at most a small fraction of the Eddington limit. However, these studies have focused only on the regime where the ionization radius is much larger than the sonic radius. Furthermore, self-gravity of the gas was ignored on the assumption that the BH's gravity dominates in the ionized zone. But even if the initial density is relatively low, as long as the radius of the gas sphere is much larger than the ionization region, self-gravity would build up the density just outside the ionization radius to a high level after one free-fall timescale. 

Here, we report new findings of accretion onto a stellar-mass BH ($\MBH = 100\, \Msun$) embedded in a primordial gas sphere from numerical simulations, which cover a wider range of numerical and physical parameters than previous studies. Our model includes not only the relevant feedback processes, but also self-gravity of the gas. We use a modified version of the public, grid-based hydrodynamics code VH-1 \citep{Blondin1993} to follow the accretion in a spherical geometry.  VH-1 is based on the piece-wise parabolic method \citep{Colella1984}, and uses a Lagrange-Remap approach to solve the Euler equations. 

The modifications we made to VH-1 include two major aspects: 
a logarithmic grid to ensure sufficiently high resolution near the central hole while covering a large region with a reasonable number of grid cells, and a unique treatment of radiative feedback by using a finely pre-computed two-parameter grid, the ionization parameter and gas temperature, to tabulate the cooling, heating, and radiation pressure. Our feedback algorithm is an improvement to those previously used \citep{Ciotti2001, Sazonov2005, Ciotti2007} in that it calculates the cooling, heating, and radiation force more accurately. Moreover, it can handle high-density clouds accurately and efficiently. These modifications allow us to achieve an unprecedentedly high spatial resolution of $10^{11}$~cm ($\sim 10^{-7}$~pc), and simulate clouds of a wide range of gas density in $10^{5} - 10^{11}\, \cm^{-3}$. In particular, the ability to reach densities several orders of magnitude higher than the previous limit of $10^{7}\, \rm{cm}^{-3}$ is critical to our new results.

The paper is organized as follows. In \S~2, we describe our analytical considerations and numerical methods in detail. In \S~3, we first demonstrate that including self-gravity enhances the gas density just outside the ionization radius by several orders of magnitude, which leads to enhanced accretion rate. We then present various simulations with higher but fixed ambient densities without including self-gravity, and show that beyond a certain critical density, the accretion rate indeed reaches Eddington limit, and that there exists a correlation between BH accretion rate and ambient gas density. We discuss the implications of our results to large scale numerical simulations of SMBH formation, and summarize in \S~4.

\section{Methodology}
\label{sec:method}

\subsection{Analytical Considerations}

In the analytical work of \citet{Milos2009B} and numerical simulations of
\citet{Milos2009A} and \citet{Park2011}, there is an implicit assumption that the
ionization radius, $\ri$, is much larger than the sonic radius, $\rs$.  
As we will show later, the conclusion that the accretion rate is significantly
suppressed by radiative feedback processes is in fact critically dependent on
this assumption, more accurately, the assumption that $\ri \gg \rB$, where
$\rB$ is the standard Bondi accretion radius ignoring radiative feedback
processes: 

\begin{equation}
\label{eq:rb}
\rB = 5\times 10^{15} \frac{M_2}{T_{0,4}}\mbox{~cm},
\end{equation}
where $M_2$ is the BH mass in unit of 100~$M_{\sun}$, and $T_{0,4}$ is
the ambient gas temperature at infinity in unit of $10^4$~K.

According to \citet{Milos2009B}, the ionization radius $\ri$ takes the form:

\begin{equation}
\label{eq:ri}
\ri = 4.7\times
10^{18}\frac{l^{1/3}f_{ion}^{1/3}M_2^{1/3}T_{\mbox{HII},4.7}}{f_{turb}^{2/3}n_5^{2/3}T_{\mbox{HI},3.7}^{2/3}}\mbox{~cm},
\end{equation}
where $f_{ion}$ is the average fraction of energy of an absorbed photon that
goes to photo-ionization, and is $1/3$ for a power-law spectrum with an index
of $1.5$, as we will assume in the present work; $l$ is the BH
luminosity relative to the Eddington luminosity for Thomson scattering; and $n_5$
is the ambient gas density in unit of $10^5$~cm$^{-3}$. 

By taking other numerical factors to be of order unity, we have $\ri<\rB$, if
$n_5M_2>2l^{1/2}\times 10^4$. For a BH with $M_2=1$, and at near
Eddington limit, this means $n > 2\times 10^9$~cm$^{-3}$. This density is much higher than those considered in the simulations of \citet{Milos2009A} and \citet{Park2011}. It is therefore not
surprising that they found strongly suppressed accretion rates. However, one
can naively imagine that there should be qualitative differences in the
properties of the accretion flow when the $\ri=\rB$ boundary is crossed. The
reason is that $\ri$ primarily determines the region where photon-heating and
photo-ionization are important processes, while $\rB$ determines the region
where the BH's gravitational field starts to significantly modify the
gas flow. When $\ri < \rB$, the radiation feedback is not important at radius
larger than $\rB$ except for Thomson scattering. The gas flow
can therefore achieve significant inward velocity of the order of local sound
speed at $\rB$. On the other hand, at locations immediately inside $\ri$,
radiation feedback may slow down the inward velocity significantly, creating a
strong shocked region between $\ri$ and $\rB$. It is this region that may help
to maintain the accretion rate of the flow at near Eddington limit.

One critical question is therefore what kind of gas density can one expect to
find near the accreting BHs. In this paper, we suggest that densities
as high as, or higher than $10^{9}$~cm$^{-3}$ might be quite common. One
reason is that the near isothermal collapse of the gas outside of $\ri$ under
self-gravity can naturally enhance the ambient density to high values. The
self-gravity of gas was ignored in the numerical simulations of
\citet{Milos2009A} and \citet{Park2011}, on the ground that the BH's gravity dominate at $r<\ri$. However, even if we start at a relatively low
density initially, as long as the radius of the gas sphere is much larger than
$\ri$, the densities just outside $\ri$ may reach very high levels shortly
after one free-fall time scale. In this work we explore this possibility and the resulting accretion rates with one-dimensional hydrodynamic simulations in spherical geometry.

\subsection{Numerical Methods}

As discussed above, there are two major feedback processes that affect the BH accretion, namely the photo-ionization heating and the radiation pressure. To study these processes and their effects on the BH accretion, one must resolve the spatial scale smaller than the sonic radius, which is of the order of $10^{14}$~cm for a 100~$\Msun$ BH.  Moreover, in order to investigate the dependence of BH accretion on gas properties, a variety of gas density must be considered.

In order to achieve a sufficiently high spatial resolution, and simulate a wide range of gas density efficiently, we use a modified version of the one-dimensional hydrodynamics code VH-1 by \cite{Blondin1993}, which is publicly available\footnote{http://wonka.physics.ncsu.edu/pub/VH-1/}. VH-1 is based on the piece-wise parabolic method \citep{Colella1984}, and uses a Lagrange-Remap approach to solve the Euler equations: 

\begin{eqnarray}
\label{eq:euler}
\partial_t\rho + \nabla\cdot(\rho\textbf{u}) &=& 0 \nonumber\\
\partial_t(\rho\textbf{u}) + \nabla\cdot(\rho\textbf{uu}) + \nabla P &=&
\rho\textbf{a} \nonumber\\
\partial_t(\rho E) + \nabla\cdot(\rho E\textbf{u} + P\textbf{u}) &=&
\rho\textbf{u}\cdot\textbf{a} + H - C,
\end{eqnarray}
where the primary variables are the mass density $\rho$, gas pressure $P$, and fluid velocity $\textbf{u}$. In our implementation, the acceleration field $\textbf{a}$ includes both gravity and radiative pressure forces, and $H$ and $C$ are heating and cooling source terms arising from radiative feedback.

The total energy per unit mass $E$ is the sum of the kinetic energy and the internal energy $e$:

\begin{equation}
E = \frac{1}{2}{\bf u}^2 + e,
\end{equation}

and the internal energy is related to the pressure through the equation of state for an ideal gas for a given adiabatic index $\gamma$.

\begin{equation}
P = \rho e (\gamma-1).
\end{equation}

The modifications we made to VH-1 include two main aspects. The first one is the straightforward change of a uniform radial grid in spherical geometry to a logarithmic one. This is to ensure sufficiently high resolution near the central BH while covering a large spherical region with a reasonable number of grid cells. The second one involves a detailed treatment of radiative feedback. Specifically, we added the heating and cooling terms in the energy conservation equation, and the radiative pressure forces in the momentum and energy conservation equations. 

The radiative feedback is modeled by formulating the cooling, heating, and radiation pressure with two parameters: the ionization parameter $\xi$ and gas temperature $T$.  We assume that the photo-ionized region around the BH can be fully described by $\xi$ and $T$. The ionization parameter $\xi$ is defined as:

\begin{equation}
\xi=\frac{L(r)}{n(r)r^2},
\end{equation}
where $L(r)$ is the local luminosity, and $n(r)$ is the hydrogen number density at radius $r$. The BH luminosity $L(0)$ is assumed to have a power law spectrum, $L_{\nu}\sim \nu^{-3/2}$, and the attenuation of the luminosity is described by 

\begin{equation}
\label{eq:rt}
\frac{d L(r)}{d r} = -4\pi r^2 \rho c a^{r}_{abs},
\end{equation}
where $a^r_{abs}$ is the radiative acceleration due to photo-absorption, and $c$ is the speed of light. In solving Equation~\ref{eq:rt}, we make another simplifying assumption that the spectral shape of $L(r)$ is the same as the original $L(0)$, so that the heating, cooling, and radiation acceleration at any radius can be obtained from pre-computed tables of the $\xi$-$T$ grid,

\begin{eqnarray}
H &=& n^2\tilde{H}(\xi, T) \nonumber\\
C &=& n^2\tilde{C}(\xi, T) \nonumber\\
a^{r}_{abs} &=& n\tilde{a}^{r}_{abs}(\xi, T)\nonumber\\
a^{r}_{line} &=& n\tilde{a}^{r}_{line}(\xi, T),
\end{eqnarray}
where the scaling on the local hydrogen density is explicitly factored out, and $a^{r}_{line}$ is the radiation acceleration mainly due to resonant $\rm {Ly\alpha}$ line scattering. 

We compute these two dimensional tables using the photo-ionization code CLOUDY version 08.00 \citep{Ferland1998}, assuming a primordial elemental abundance, and a power-law input spectrum with index $-1.5$ in the energy range of 13.6~eV to 100~keV. In the hydrodynamics update of the momentum and energy conservation equations, the heating, cooling, and radiation forces are then interpolated from these tables, in conjunction with the solution of Equation~\ref{eq:rt}. 

In addition to photo-absorption and resonant line scattering, Thomson scattering also contributes to the radiation pressure force, which is treated separately as $a^r_{cont} = -a^g(r)L(0)/L_{edd}(r)$, where $a^g(r)$ and $L_{edd}$ are the gravitational acceleration and Eddington luminosity of the BH, respectively. We note that the unattenuated BH luminosity, $L(0)$, is used in calculating $a^r_{cont}$, as the absorbed luminosity is assumed to be re-emitted at energies below the ionization threshold, which also contributes to Thomson scattering.

Our feedback method is an improvement to those previously used (e.g., \citealt{Ciotti2001, Sazonov2005, Ciotti2007}) in that it computes the heating, cooling, and radiation pressure forces more accurately, and applies to a temperature range of $1 \le T \le 10^8$~K, much larger than the previous range $10^4 \le T \le 3\times 10^7$~K. As demonstrated in the main page, our simulations produce similar results as those by \cite{Milos2009A} and \cite{Park2011} using more sophisticated treatments on the same problems. Moreover, this method allows us to simulate high-density clouds accurately and efficiently, which is critical in our investigation.

Overall, these modifications allow us to achieve an unprecedentedly high spatial resolution of $10^{11}$~cm ($10^{-7}$~pc), and model a large range of gas density $10^{5} - 10^{11}\, \cm^{-3}$. In particular, the ability to reach densities several orders of magnitude higher than the limit of $10^{7}\, \rm{cm}^{-3}$ in previous work by \cite{Milos2009A} and \cite{Park2011} is the key to our new findings. 

\begin{deluxetable*}{lllccccccc}[h]
\tablecaption{\label{tab:sims}Important Parameters and Properties of Simulations Performed}
\tablehead{
{Runs} &
{$n_0$~(cm$^{-3}$)\tablenotemark{[1]}} &
{$r_{min}$~(pc)\tablenotemark{[2]}} &
{$r_{max}$~(pc)\tablenotemark{[3]}} &
{$N_{grid}$\tablenotemark{[4]}} &
{$\dot{M}_{min}$\tablenotemark{[5]}} &
{$\dot{M}_{avg}$\tablenotemark{[6]}} &
{$\tau_{cycle}$~(yr)\tablenotemark{[7]}} &
{$\ri$~($10^{-3}$~pc)\tablenotemark{[8]}}
}
\startdata
A & $10^7$ & $5\times 10^{-5}$ & 1 & 1200 & 0.15 & 0.49 & 200 & 6.8 \\
B\tablenotemark{[9]} & $10^7$ & $5\times 10^{-5}$ & 1 & 1200 & $\cdots$ & $\cdots$ & $\cdots$ & $\cdots$ \\
C & $10^5$ & $10^{-4}$ & 1 & 1000 & 0.007 & 0.031 & 1500 & 40 \\
D & $5\times 10^5$ & $10^{-4}$ & 1 & 1000 & 0.01 & 0.099 & 720 & 25 \\
E & $10^6$ & $10^{-4}$ & 1 & 1000 & 0.02 & 0.15 & 538 & 18 \\
F & $5\times 10^6$ & $10^{-4}$ & 1 & 1000 & 0.06 & 0.40 & 295 & 11 \\
G & $2.5\times 10^7$ & $5\times 10^{-5}$ & 1 & 1200 & 0.3 & 0.70 & 118 & 4.6 \\
H & $5\times 10^7$ & $5\times 10^{-5}$ & 1 & 1200 & 0.6 & 0.91 & 76 & 3.1 \\
I & $10^8$ & $5\times 10^{-5}$ & 1 & 1200 & 0.85 & 1.13 & 36 & 2.2 \\
J & $5\times 10^8$ & $10^{-6}$ & 1 & 1200 & 1.5 & 1.61 & 33 & 0.72 \\
K & $10^{9}$ & $10^{-6}$ & 1 & 1200 & 1.72 & 1.72 & 21 & 0.47 \\
L & $10^{10}$ & $5\times 10^{-7}$ & 1 & 1200 & 1.82 & 1.82 & 12 & 0.17 \\
M\tablenotemark{[10]} & $10^{11}$ & $10^{-7}$ & $10^{-2}$ & 2000 & 1.94 & 1.94 & 2.0 & 0.057 \\
\enddata

\tablenotetext{[1]}{Ambient density of the gas sphere.} 
\tablenotetext{[2]}{Inner radius of the simulation volume.} 
\tablenotetext{[3]}{Outer radius of the simulation volume.} 
\tablenotetext{[4]}{Number of points for the logarithmic radial grid.}
\tablenotetext{[5]}{Asymptotic minimum accretion rate in unit of $10^{-6}\, \Msun\, \yr^{-1}$.} 
\tablenotetext{[6]}{Asymptotic average accretion rate in unit of $10^{-6}\, \Msun\, \yr^{-1}$.}
\tablenotetext{[7]}{Asymptotic period of the oscillation cycle.} 
\tablenotetext{[8]}{Ionization radius where the ionization fraction is 0.5.} 
\tablenotetext{[9]}{Only run B includes self-gravity. All other runs are without self-gravity in order to exam the main dependence of the accretion process on the ambient density.} 
\tablenotetext{[10]}{At $n_0=10^{11}$~cm$^{-3}$, the outer radius is reduced to  0.01 pc to accommodate the much smaller inner radius, and to make it computationally feasible.} 
\end{deluxetable*}

\subsection{Initial Conditions and Parameters of Simulations}

In this study, we model the radial accretion onto a central BH embedded in a gas sphere. We assume a BH mass $\MBH = 100\, \Msun$, a radiative efficiency of 0.1, and the gas clouds are assumed to be metal-free and uniform with an ambient density $n_0$, and an initial temperature $T = 10^4$~K. Outflow and inflow boundary conditions are used in the inner and outer boundaries, respectively.  The inner and outer boundaries of the sphere are chosen in such a way that both sonic radius and ionization radius are located well within the simulation region in order to sufficiently resolve the accretion flow. 

A total of 13 simulations with different initial hydrogen densities were performed in the present work. The most important parameters and properties of these models are listed in Table~1. In this table, column 1 lists the name of the simulations. Columns 2-5 give the numerical parameters: the ambient density of the gas cloud, the inner and outer radius of the simulated sphere, and number of grid points, respectively. Columns 6-9 give the properties of the accretion process: the minimum and average accretion rates, the oscillation period of the accretion process, and the ionization radius where the ionization fraction is 0.5, respectively.

Run A used the same initial conditions as in other previous work by \cite{Milos2009A} and \cite{Park2011}. It produced similar results of many aspects of the accretion process, including the general intermittent pattern, accretion rates, and oscillation period. This confirms the previous findings that radiative feedback strongly suppresses BH accretion. In Run B, self-gravity of the gas was included. We find that self-gravity greatly modifies the accretion flow. It helps to maintain a large inflow rate and increase density buildup outside of the ionization radius $\ri$, which leads to the shrink of $\ri$. When the density reaches a critical threshold, the radiative feedback becomes less effective, so the accretion rate can reach Eddington limit. Runs C -- M covered a wide range of gas density of $10^{5 - 11}\, \cm^{-3}$ in order to explore the dependence of accretion on gas density. In these simulations, the self-gravity is deliberately turned off to avoid the density enhancement due to the collapse outside of $\ri$. Such controlled simulations are not only more computationally tractable than simply allowing the gas to collapse infinitely under self-gravity, they also better illustrate the main dependence of the accretion process on the ambient density.

\section{Results}
\label{sec:results}

\subsection{Enhanced Accretion by Self-gravity of Gas}

\begin{figure}
\begin{center}
\includegraphics[width=3.3in]{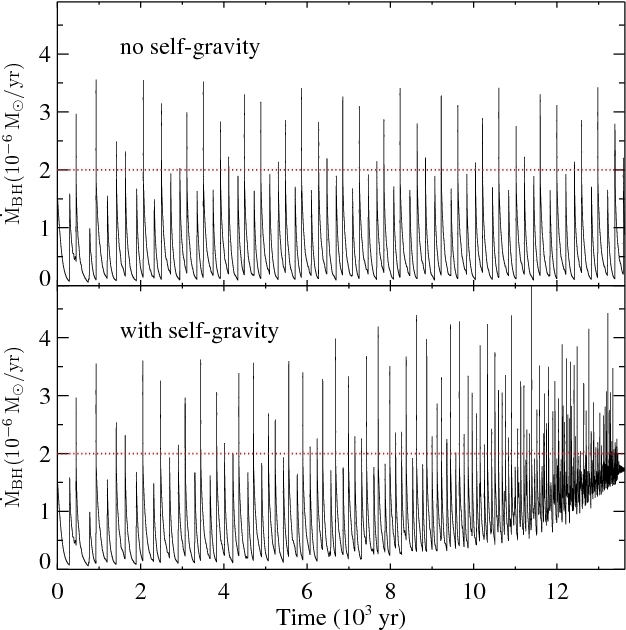} 
\caption{Time evolution of accretion rates onto a $100\, \Msun$ black hole embedded in a uniform, primordial gas sphere with ambient density of $10^7\, \cm^{-3}$, {\it without} (top panel) and {\it with} (bottom panel) self-gravity of the gas in the simulation. The dotted line in each panel indicates the Eddington rate $\Mdotedd \sim 2\times 10^{-6}\, \Msun\, \yr^{-1}$. The accretion behavior exhibits a periodical oscillation caused by the alternating inflow driven by gravity and external pressure and outflow driven by radiative feedback from the BH accretion. Without gas self-gravity, the average accretion is about 25\% of the Eddington rate, but with self-gravity, it reaches $\sim$~86\% of the Eddington rate after one free-fall timescale of the gas sphere.}
\label{fig_acc_t}
\end{center}
\end{figure}

The BH accretion rate is dictated by the interplay between inflow driven by gravity and external pressure and outflow driven by radiation pressure. Without gas self-gravity, the accretion exhibits a pattern of periodic oscillations, as demonstrated in Figure~\ref{fig_acc_t} ({\it top panel}) for an ambient density of $10^7\, \cm^{-3}$. The intermittency is due to alternating expulsion and fallback of the gas flow with an average period of approximately 200 years. The maximum rates occasionally exceed the Eddington limit ($\Mdotedd \sim 2\times 10^{-6}\, \Msun\, \yr^{-1}$), but the mean remains roughly constant at about 25\% of the Eddington rate over a very long timescale. These results are in good agreement with previous ones from 2-dimensional simulations \citep{Milos2009A, Park2011}. Similar intermittent behavior were also seen in other simulations of Bondi-like (e.g., \citealt{Krumholz2005, Ciotti2007}), or rotating accretion (e.g., \citealt{Proga2008}) on different scales.

However, when self-gravity of the gas is included, the BH accretion pattern changes dramatically, as shown in Figure~\ref{fig_acc_t} ({\it bottom panel}). It is clear that the minimum accretion rate within one oscillation cycle gradually increases. In about one free-fall timescale of the gas sphere ($\sim 1.35\times 10^4$ years), the mean accretion rate reaches a constant at $\sim 1.72\times 10^{-6}\, \Msun \yr^{-1}$, or about 86\% of the Eddington rate. 

The reason for such a significant enhancement is that the BH accretion rate critically depends on the relation between ionization radius $\ri$ and sonic radius, or more accurately, the standard Bondi accretion radius ignoring radiative feedback $\rB$ ($\sim 5\times 10^{15}\, \cm$ for a $100\, \Msun$ BH in a $10^4$~K cloud):  the accretion is strongly suppressed by radiative feedback if $\ri \gg \rB$, but is less affected if $\ri < \rB$. The properties of the accretion flow change when the $\ri=\rB$ boundary is crossed. At radius smaller than $\ri$, radiative heating and photo-ionization dominate, while the BH's gravitational field starts to substantially modify the gas flow at radius close to $\rB$. When $\ri < \rB$, the radiation feedback is not important at radius larger than $\rB$ except for Thomson scattering. The gas flow can therefore achieve a large inward velocity of the order of local sound speed at $\rB$. On the other hand, when $\ri > \rB$, radiation feedback may reduce the local accretion rate significantly, creating a strong shocked region between $\ri$ and $\rB$. It is this region with enhanced density that helps to maintain a relatively large inflow rate only limited by Thomson scattering. Self-gravity gradually overcomes the radiation force and increases density buildup outside of $\ri$, which leads to the shrinking of $\ri$ to within $\rB$. In the regime where $\ri < \rB$, the accretion rate can approach to Eddington limit when the density at $\rB$ reaches a critical value.

\subsection{Dependence of Accretion on Ambient Gas Density}

\begin{figure}
\begin{center}
\includegraphics[width=3.3in]{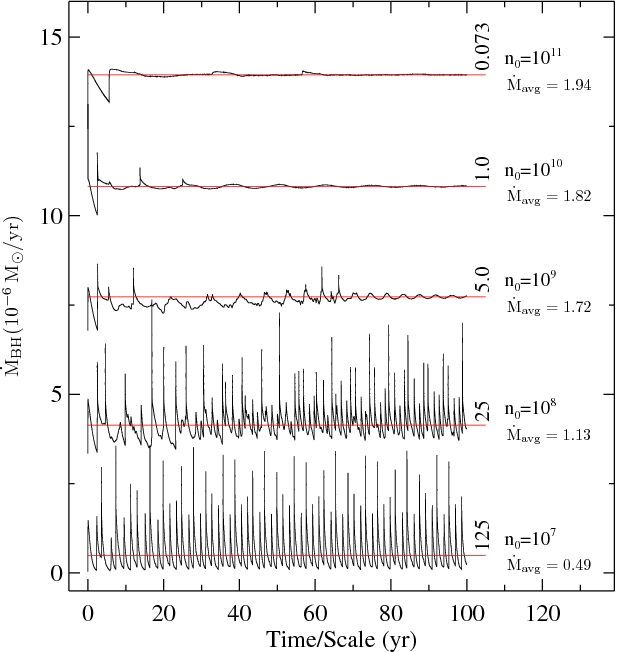} 
\caption{Time evolution of BH accretion rate for different ambient gas density, $n_0=10^{7}$, $10^{8}$, $10^{9}$, $10^{10}$, and $10^{11}$~cm$^{-3}$, respectively. Note the accretion curves of the last four densities are shifted vertically for easy comparison, and the red line represents the mean accretion rate, $\dot{\rm M}_{\rm avg}$, of each simulation with the actual value (in unit of $10^{-6}\, \Msun \yr^{-1}$) given next to it. The horizontal axis is the scaled time, where the scale factor for different density is given in vertical orientation at the end of the corresponding accretion curve. In these simulations, self-gravity of the gas is deliberately turned off to avoid density enhancement due to the collapse outside of $\ri$. Such controlled simulations not only better unveil the important dependence of the accretion process on the ambient density, they are also more computationally tractable than simply allowing the gas to collapse infinitely under self-gravity.}  
\label{fig_acc_tn}
\end{center}
\end{figure}

\begin{figure}
\begin{center}
\includegraphics[width=3.3in]{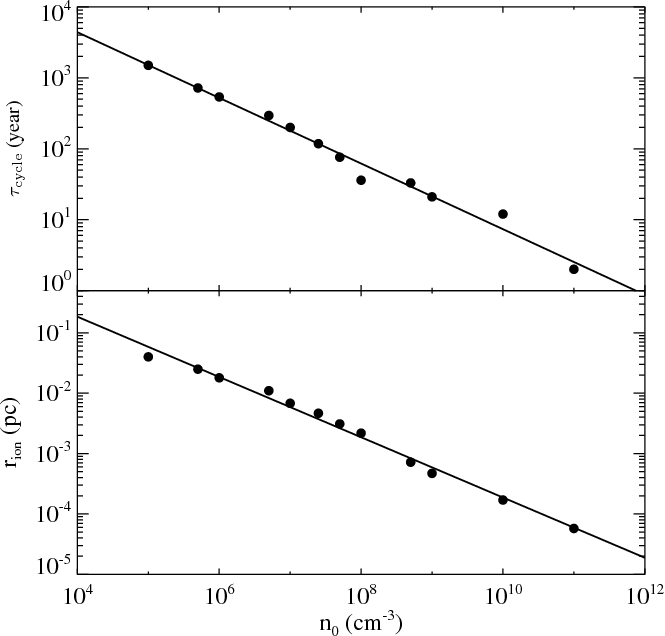}
\caption{Dependence of the oscillation period of the BH accretion, $\tau_{\rm cycle}$ ({\it top panel}), and ionization radius $\ri$ ({\it bottom panel}), on the ambient density, $n_0$, of the gas sphere. The filled circles are simulation data, while the solid lines are fitting curves. Both $\tau_{\rm cycle}$ and $\ri$ decrease with density: $\tau_{\rm cycle} = 3\times10^5(n_0/\cm^{-3})^{-0.46}$ years, and $\ri = 18.3(n_0/\cm^{-3})^{-0.5}$ pc. At $n_0 \gtrsim 10^9\, \cm^{-3}$, the ionization radius becomes $\ri < \rB$ ($\rB \sim 1.6\times 10^{-3}\, \pc$), the radiation feedback is weakened, and the oscillation is strongly damped. } 
\label{fig_tri_n}
\end{center}
\end{figure}

An important question is then what the critical density would be. Figure~\ref{fig_acc_tn} shows the time evolution of accretion rate at several different densities. As the density increases, the accretion rate becomes constant and approaches to the Eddington limit. 

The oscillation period also decreases as density increases, and scales as $n_0^{-0.46}$, as illustrated in Figure~\ref{fig_tri_n} ({\it top panel}). When the density reaches $\sim 10^{9}\, \cm^{-3}$, the oscillation is significantly damped after an initial time period, and the ionization radius $\ri$ drops to below $\rB$, as shown in Figure~\ref{fig_tri_n} ({\it bottom panel}). The radiation feedback becomes less effective. As a result, the mean asymptotic accretion rate rises to about 90\% of the Eddington limit. A density of $10^{9}\, \cm^{-3}$ or higher may be common in collapsing clouds in massive dark matter halos in the early universe, as reported in high-resolution simulations of the first protostars (e.g., \citealt{Yoshida2008}). This density is much higher than those considered in previous work of similar problem (e.g., \citealt{Alvarez2009, Milos2009A, Park2011}). It is therefore not surprising that they found severely reduced accretion rates.  

\subsection{A Universal Correlation Between Black Hole Accretion Rate and Ambient Gas Density}

\begin{figure}
\begin{center}
\includegraphics[width=3.3in]{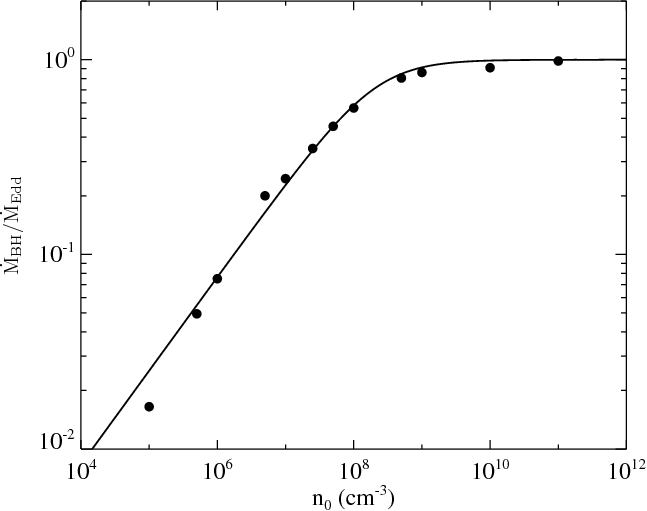}
\caption{Correlation between the BH asymptotic accretion rate and the ambient gas density. The filled circles represent data from simulations with radiation feedback, while the solid line is the least-square fitting, which takes the following form: $\Mdotbh=\Mdotedd\left(\frac{n_0/n_{\rm c}}{1+n_0/n_{\rm c}}\right)^\beta$, where $n_{\rm c} = 2\times 10^8\, \cm^{-3}$, and $\beta=0.48$. This plot shows that, at densities below $2\times 10^8\, \cm^{-3}$, radiative feedback dominates and strongly reduces the accretion to be sub-Eddington, while above that, radiative feedback becomes less effective, and the accretion reaches the Eddington rate. }     
\label{fig_acc_n}
\end{center}
\end{figure}

Figure~\ref{fig_acc_n} shows the relationship between the BH accretion rate and ambient density, which can be described as $\Mdotbh=\Mdotedd\left(\frac{n_0/n_{\rm c}}{1+n_0/n_{\rm c}}\right)^\beta$, where $n_{\rm c}$ is the characteristic density at which the asymptotic accretion rate approaches to the Eddington limit. As shown in the Figure, the fitting of the simulation data gives $n_{\rm c} = 2\times 10^8\, \cm^{-3}$, and $\beta=0.48$. At densities above $2\times 10^8\, \cm^{-3}$, radiative feedback other than Thomson scattering becomes less effective, and the accretion reaches the Eddington limit, while below that, radiative heating and photo-ionization feedback dominates and strongly suppresses the accretion. This relationship would imply an accretion rate larger than the Bondi rate at densities below $\sim 10^2\, \cm^{-3}$. However, because both ionization and heating timescales decrease as density becomes smaller, they eventually become much less than the dynamical timescale of the gas. Under such conditions, the radiative feedback plays a negligible role, and the accretion becomes Bondi-like.

More intriguingly, we repeat the simulations for a BH mass range of $10^2 - 10^9\, \Msun$, and 
find that this correlation is universal over a wide range of BH mass and gas density, with a simple scaling relation between the BH mass and the critical density, $n_{\rm crit}\MBH=({2\times 10^8}/{\cm^{-3}})({10^2}/{\Msun})$. This scaling relation is easy to understand, as the hydrodynamic equations governing the accretion process without self-gravity can be shown to depend only on $n_0\MBH$, under the assumption that the density is sufficiently high so that the local gas temperature is close to the thermal equilibrium determined by the heating and cooling functions. That means, for a given BH mass, there exists a critical density, above which the accretion rate can reach the Eddington limit. 

This general relation between BH accretion and ambient gas density has important implications and applications in studies of BH growth. We note that Bondi accretion has been commonly used in simulations of BH growth (e.g., \citealt{Li2007A, Johnson2007, Alvarez2009, DiMatteo2005, Springel2005B, Hopkins2006A, DiMatteo2008, Sijacki2009, DiMatteo2011}). This simplified prescription neglects the effects of radiation feedback and overestimates the accretion rate by up to two orders of magnitude at some densities below $10^8\, \cm^{-3}$. Our results and the fitting formula above can serve as a more realistic recipe for BH accretion, and can be implemented directly into numerical simulations. 

\section{Summary}

To summarize, we have presented a set of one-dimensional hydrodynamic simulations of the accretion of a black hole embedded in a primordial gas cloud, using the modified grid-based VH-1 code. We include not only important feedback processes from the accreting black hole, but also self-gravity of the gas. We achieved an unprecedentedly high spatial resolution of $10^{11}$ cm, and covered a wide range of gas density of $10^{5} - 10^{11}\, \cm^{-3}$. These advantages allowed us to study the accretion process in regimes not explored by previous work, and unveil the following new findings: 

\begin{enumerate}

\item The accretion behavior exhibits a periodical oscillation caused by the alternating inflow driven by gravity and external pressure and outflow driven by radiative feedback from the BH accretion. Self-gravity of the gas can boost the accretion rate by building up high gas density which weakens the feedback effects.  Without gas self-gravity, the average accretion is about 25\% of the Eddington rate, but with self-gravity, it reaches $\sim$~86\% of the Eddington rate after one free-fall timescale of the gas sphere.

\item The accretion depends strongly on the ambient gas density. For a given black hole mass, there exists a critical density above which the accretion can reach Eddington limit. For example, for a 100~$\Msun$ BH, the critical density is $\sim 10^{9}\, \cm^{-3}$.

\item There exists a universal correlation between black hole accretion rate and ambient gas density: $\Mdotbh=\Mdotedd\left(\frac{n_0/n_{\rm c}}{1+n_0/n_{\rm c}}\right)^\beta$, where $n_{\rm c} = 2\times 10^8\, \cm^{-3}$, and $\beta=0.48$. This fitting formula may serve as a realistic recipe for BH accretion, and can be implemented directly into numerical simulations.  

\end{enumerate}

In the rare, high-density peaks ($5-6\, \sigma$) of the cosmic filaments where the first most massive dark matter halos collapsed, the first stars were born and might evolve into the first stellar-mass BHs. The extremely deep gravitational potential in these regions retained an abundant gas supply, and induced vigorous interactions and mergers of protogalaxies. Strong gravitational torques removed angular momentum from the highly shocked and dense gas and transported it to fuel the central BHs, while triggered global starburst on a larger scale. Seed BHs in such a gas-rich and dynamical environment may grow rapidly and co-evally with host galaxies to become the first luminous quasars at the cosmic dawn.

\section{Acknowledgments}
I thank Tom Abel, Tiziana Di Matteo, Mike Eracleous, Carlos Frenk, Alex Heger, Lars Hernquist, Peter M{\'e}sz{\'a}ros, Peng Oh, Massimo Ricotti, and Daniel Schaerer for stimulating discussions and helpful comments. Support from NSF grants AST-0965694 and AST-1009867 is gratefully acknowledged. I thank the Institute for Theory and Computation (ITC) at Harvard University where the project was started for warm hospitality, and the Research Computing and Cyberinfrastructure unit of Information Technology Services at The Pennsylvania State University for providing computational resources and services that have contributed to the research results reported in this paper. URL: http://rcc.its.psu.edu. 

\bibliography{ref}

\end{document}